\documentclass[12pt,a4paper]{article}

\usepackage{ifthen} 
\newboolean{pdflatex}
\setboolean{pdflatex}{true} 

\newboolean{articletitles}
\setboolean{articletitles}{true} 

\newboolean{uprightparticles}
\setboolean{uprightparticles}{false} 

\newboolean{inbibliography}
\setboolean{inbibliography}{false} 

\usepackage{longtable} 
\usepackage{subfigure}

\usepackage{microtype}
\usepackage{lineno}  
\usepackage{xspace} 
\usepackage{caption} 

\usepackage{graphicx}  
\usepackage{color}
\usepackage{colortbl}
\graphicspath{{./figs/}} 

\usepackage{amsmath} 
\usepackage{amssymb}
\usepackage{amsfonts}
\usepackage{upgreek} 

\newcommand*\patchAmsMathEnvironmentForLineno[1]{%
\expandafter\let\csname old#1\expandafter\endcsname\csname #1\endcsname
\expandafter\let\csname oldend#1\expandafter\endcsname\csname
end#1\endcsname
 \renewenvironment{#1}%
   {\linenomath\csname old#1\endcsname}%
   {\csname oldend#1\endcsname\endlinenomath}%
}
\newcommand*\patchBothAmsMathEnvironmentsForLineno[1]{%
  \patchAmsMathEnvironmentForLineno{#1}%
  \patchAmsMathEnvironmentForLineno{#1*}%
}
\AtBeginDocument{%
\patchBothAmsMathEnvironmentsForLineno{equation}%
\patchBothAmsMathEnvironmentsForLineno{align}%
\patchBothAmsMathEnvironmentsForLineno{flalign}%
\patchBothAmsMathEnvironmentsForLineno{alignat}%
\patchBothAmsMathEnvironmentsForLineno{gather}%
\patchBothAmsMathEnvironmentsForLineno{multline}%
}

\usepackage{hyperref}    
\usepackage[all]{hypcap} 

\def\lhcb {\mbox{LHCb}\xspace}

\ifthenelse{\boolean{uprightparticles}}%
{

 \def\Pmu         {\ensuremath{\upmu}\xspace}                 
 \def\Pnu         {\ensuremath{\upnu}\xspace}                 
                  
 \def\Ppi         {\ensuremath{\uppi}\xspace}

 \def\Pchi        {\ensuremath{\upchi}\xspace}                 
 \def\Ppsi        {\ensuremath{\uppsi}\xspace}

 \def\PDelta      {\ensuremath{\Delta}\xspace}                 
 \def\PXi      {\ensuremath{\Xi}\xspace}                 
 \def\PLambda      {\ensuremath{\Lambda}\xspace}                 
 \def\PSigma      {\ensuremath{\Sigma}\xspace}                 
 \def\POmega      {\ensuremath{\Omega}\xspace}                 
 \def\PUpsilon      {\ensuremath{\Upsilon}\xspace}                 
 
 \def\PB      {\ensuremath{\mathrm{B}}\xspace}                 
                  
 \def\PD      {\ensuremath{\mathrm{D}}\xspace}

 \def\PJ      {\ensuremath{\mathrm{J}}\xspace}                 
 \def\PK      {\ensuremath{\mathrm{K}}\xspace}

 \def\Pb      {\ensuremath{\mathrm{b}}\xspace}                 
 \def\Pc      {\ensuremath{\mathrm{c}}\xspace}                 
                  
 \def\Pe      {\ensuremath{\mathrm{e}}\xspace}

 \def\Pi      {\ensuremath{\mathrm{i}}\xspace}

 \def\Pp      {\ensuremath{\mathrm{p}}\xspace}

 \def\Ps      {\ensuremath{\mathrm{s}}\xspace}

}
{

 \def\Pmu         {\ensuremath{\mu}\xspace}                 
 \def\Pnu         {\ensuremath{\nu}\xspace}                 
                  
 \def\Ppi         {\ensuremath{\pi}\xspace}

 \def\Pchi        {\ensuremath{\chi}\xspace}                 
 \def\Ppsi        {\ensuremath{\psi}\xspace}                 
                  
 \mathchardef\PDelta="7101
 \mathchardef\PXi="7104
 \mathchardef\PLambda="7103
 \mathchardef\PSigma="7106
 \mathchardef\POmega="710A
 \mathchardef\PUpsilon="7107
                  
 \def\PB      {\ensuremath{B}\xspace}                 
                  
 \def\PD      {\ensuremath{D}\xspace}

 \def\PJ      {\ensuremath{J}\xspace}                 
 \def\PK      {\ensuremath{K}\xspace}

 \def\Pb      {\ensuremath{b}\xspace}                 
 \def\Pc      {\ensuremath{c}\xspace}                 
                  
 \def\Pe      {\ensuremath{e}\xspace}

 \def\Pi      {\ensuremath{i}\xspace}

 \def\Pp      {\ensuremath{p}\xspace}

 \def\Ps      {\ensuremath{s}\xspace}

}

\def\epem       {\ensuremath{\Pe^+\Pe^-}\xspace}

\def\mup        {\ensuremath{\Pmu^+}\xspace}
\def\mun        {\ensuremath{\Pmu^-}\xspace} 
\def\mumu       {\ensuremath{\Pmu^+\Pmu^-}\xspace}

\def\lepton     {\ensuremath{\ell}\xspace}
\def\ellm       {\ensuremath{\ell^-}\xspace}
\def\ellp       {\ensuremath{\ell^+}\xspace}

\def\neu        {\ensuremath{\Pnu}\xspace}
\def\neub       {\ensuremath{\overline{\Pnu}}\xspace}
\def\neul       {\ensuremath{\neu_\ell}\xspace}
\def\neulb      {\ensuremath{\neub_\ell}\xspace}

\def\squark    {\ensuremath{\Ps}\xspace}

\def\cquark    {\ensuremath{\Pc}\xspace}
\def\cquarkbar {\ensuremath{\overline \cquark}\xspace}
\def\ccbar     {\ensuremath{\cquark\cquarkbar}\xspace}
\def\bquark    {\ensuremath{\Pb}\xspace}

\def\pion  {\ensuremath{\Ppi}\xspace}

\def\pip   {\ensuremath{\pion^+}\xspace}
\def\pim   {\ensuremath{\pion^-}\xspace}

\def\kaon  {\ensuremath{\PK}\xspace}
  \def\Kbar  {\kern 0.2em\overline{\kern -0.2em \PK}{}\xspace}

\def\Kp    {\ensuremath{\kaon^+}\xspace}

\def\Kstarz  {\ensuremath{\kaon^{*0}}\xspace}

  \def\Dbar    {\kern 0.2em\overline{\kern -0.2em \PD}{}\xspace}

\def\Dzb     {\ensuremath{\Dbar^0}\xspace}

\def\B       {\ensuremath{\PB}\xspace}
\def\Bbar    {\ensuremath{\kern 0.18em\overline{\kern -0.18em \PB}{}}\xspace}

\def\Bu      {\ensuremath{\B^+}\xspace}

\def\Bp      {\ensuremath{\Bu}\xspace}

\def\Bd      {\ensuremath{\B^0}\xspace}
\def\Bs      {\ensuremath{\B^0_\squark}\xspace}

\def\jpsi     {\ensuremath{{\PJ\mskip -3mu/\mskip -2mu\Ppsi\mskip 2mu}}\xspace}
\def\psitwos  {\ensuremath{\Ppsi{(2S)}}\xspace}

  \def\Y#1S{\ensuremath{\PUpsilon{(#1S)}}\xspace}

\def\chic  {\ensuremath{\Pchi_{c}}\xspace}

\def\proton      {\ensuremath{\Pp}\xspace}

\def\Lz {\ensuremath{\PLambda}\xspace}
\def\Lbar {\ensuremath{\kern 0.1em\overline{\kern -0.1em\PLambda}}\xspace}

\def\Lb      {\ensuremath{\Lz^0_\bquark}\xspace}

\def\BF         {{\ensuremath{\cal B}\xspace}}

\def\BR         {\BF}
\newcommand{\decay}[2]{\ensuremath{#1\!\to #2}\xspace}         

\def\to                 {\ensuremath{\rightarrow}\xspace}

\def\qsq       {\ensuremath{q^2}\xspace}

\def\BdToKstmm    {\decay{\Bd}{\Kstarz\mup\mun}}

\def\BdKstee  {\decay{\Bd}{\Kstarz\epem}}

\def\AT#1     {\ensuremath{A_{\mathrm{T}}^{#1}}\xspace}           

\def\C#1      {\ensuremath{\mathcal{C}_{#1}}\xspace}                       
\def\Cp#1     {\ensuremath{\mathcal{C}_{#1}^{'}}\xspace}                    
\def\Ceff#1   {\ensuremath{\mathcal{C}_{#1}^{\mathrm{(eff)}}}\xspace}        
\def\Cpeff#1  {\ensuremath{\mathcal{C}_{#1}^{'\mathrm{(eff)}}}\xspace}       
\def\Ope#1    {\ensuremath{\mathcal{O}_{#1}}\xspace}                       
\def\Opep#1   {\ensuremath{\mathcal{O}_{#1}^{'}}\xspace}                    

\newcommand{\tev}{\ifthenelse{\boolean{inbibliography}}{\ensuremath{~T\kern -0.05em eV}\xspace}{\ensuremath{\mathrm{\,Te\kern -0.1em V}}\xspace}}
\newcommand{\gev}{\ensuremath{\mathrm{\,Ge\kern -0.1em V}}\xspace}
\newcommand{\mev}{\ensuremath{\mathrm{\,Me\kern -0.1em V}}\xspace}
\newcommand{\kev}{\ensuremath{\mathrm{\,ke\kern -0.1em V}}\xspace}
\newcommand{\ev}{\ensuremath{\mathrm{\,e\kern -0.1em V}}\xspace}
\newcommand{\gevc}{\ensuremath{{\mathrm{\,Ge\kern -0.1em V\!/}c}}\xspace}
\newcommand{\mevc}{\ensuremath{{\mathrm{\,Me\kern -0.1em V\!/}c}}\xspace}
\newcommand{\gevcc}{\ensuremath{{\mathrm{\,Ge\kern -0.1em V\!/}c^2}}\xspace}
\newcommand{\gevgevcccc}{\ensuremath{{\mathrm{\,Ge\kern -0.1em V^2\!/}c^4}}\xspace}
\newcommand{\mevcc}{\ensuremath{{\mathrm{\,Me\kern -0.1em V\!/}c^2}}\xspace}

\def\invfb   {\ensuremath{\mbox{\,fb}^{-1}}\xspace}

\newcommand{\stat}{\ensuremath{\mathrm{\,(stat)}}\xspace}
\newcommand{\syst}{\ensuremath{\mathrm{\,(syst)}}\xspace}

\def\deriv {\ensuremath{\mathrm{d}}}

\def\gsim{{~\raise.15em\hbox{$>$}\kern-.85em
          \lower.35em\hbox{$\sim$}~}\xspace}
\def\lsim{{~\raise.15em\hbox{$<$}\kern-.85em
          \lower.35em\hbox{$\sim$}~}\xspace}

\def\pt         {\mbox{$p_{\rm T}$}\xspace}

\def\gauss      {\mbox{\textsc{Gauss}}\xspace}

\def\tell1  {TELL1\xspace}
\def\ukl1   {UKL1\xspace}

\newcommand{\eg}{\mbox{\itshape e.g.}\xspace}

\usepackage{cite} 
\usepackage{mciteplus}

\def\ellell     {\ensuremath{\ell^+ \ell^-}\xspace}

\def\dqsq       {{\deriv q^2}\xspace}

\def\varphi    {\ensuremath{\phi}\xspace}

\def\gevgevcccc    {\ensuremath{\gev^{2} / c^{4} }\xspace}

\def\RK         {\ensuremath{R_{\kaon}}\xspace}

\def\Kll  {\ensuremath{\Kp\ellell}\xspace}
\def\Kee  {\ensuremath{\Kp\epem}\xspace}
\def\Kmm  {\ensuremath{\Kp\mumu}\xspace}
\def\JpsiKee  {\ensuremath{\jpsi(\epem)\Kp}\xspace}
\def\JpsiKmm  {\ensuremath{\jpsi(\mumu)\Kp}\xspace}

\def\BuJpsiK  {\decay{\Bu}{\jpsi\Kp}}
\def\BuJpsiKll  {\decay{\Bu}{\jpsi(\to\ellell)\Kp}}
\def\BuPsiKll  {\decay{\Bu}{\psitwos(\to\ellell)\Kp}}
\def\BuKll  {\decay{\Bu}{\Kp\ellell}}
\def\BuJpsiKmm  {\decay{\Bu}{\jpsi(\to\mumu)\Kp}}

\def\BuKmm  {\decay{\Bu}{\Kp\mumu}}
\def\BuJpsiKee  {\decay{\Bu}{\jpsi(\to\epem)\Kp}}

\def\BuKee  {\decay{\Bu}{\Kp\epem}}

\def\BdJpsiKst  {\decay{\Bd}{\jpsi\Kstarz}}

\def\BdJpsiKstmm  {\decay{\Bd}{\jpsi(\to\mumu)\Kstarz}}

\def\BdKstmm  {\decay{\Bd}{\Kstarz\mumu}}

\def\BdKstee  {\decay{\Bd}{\Kstarz\epem}}

\def\BuJpsiXee {\decay{\Bu}{(\decay{\jpsi}{\epem})\Kp X}\xspace}

\begin{document}

\renewcommand{\thefootnote}{\fnsymbol{footnote}}
\setcounter{footnote}{1}


\begin{titlepage}
\pagenumbering{roman}

\vspace*{-1.5cm}
\centerline{\large EUROPEAN ORGANIZATION FOR NUCLEAR RESEARCH (CERN)}
\vspace*{1.5cm}
\hspace*{-0.5cm}
\begin{tabular*}{\linewidth}{lc@{\extracolsep{\fill}}r}
\ifthenelse{\boolean{pdflatex}}
{\vspace*{-2.7cm}\mbox{\!\!\!\includegraphics[width=.14\textwidth]{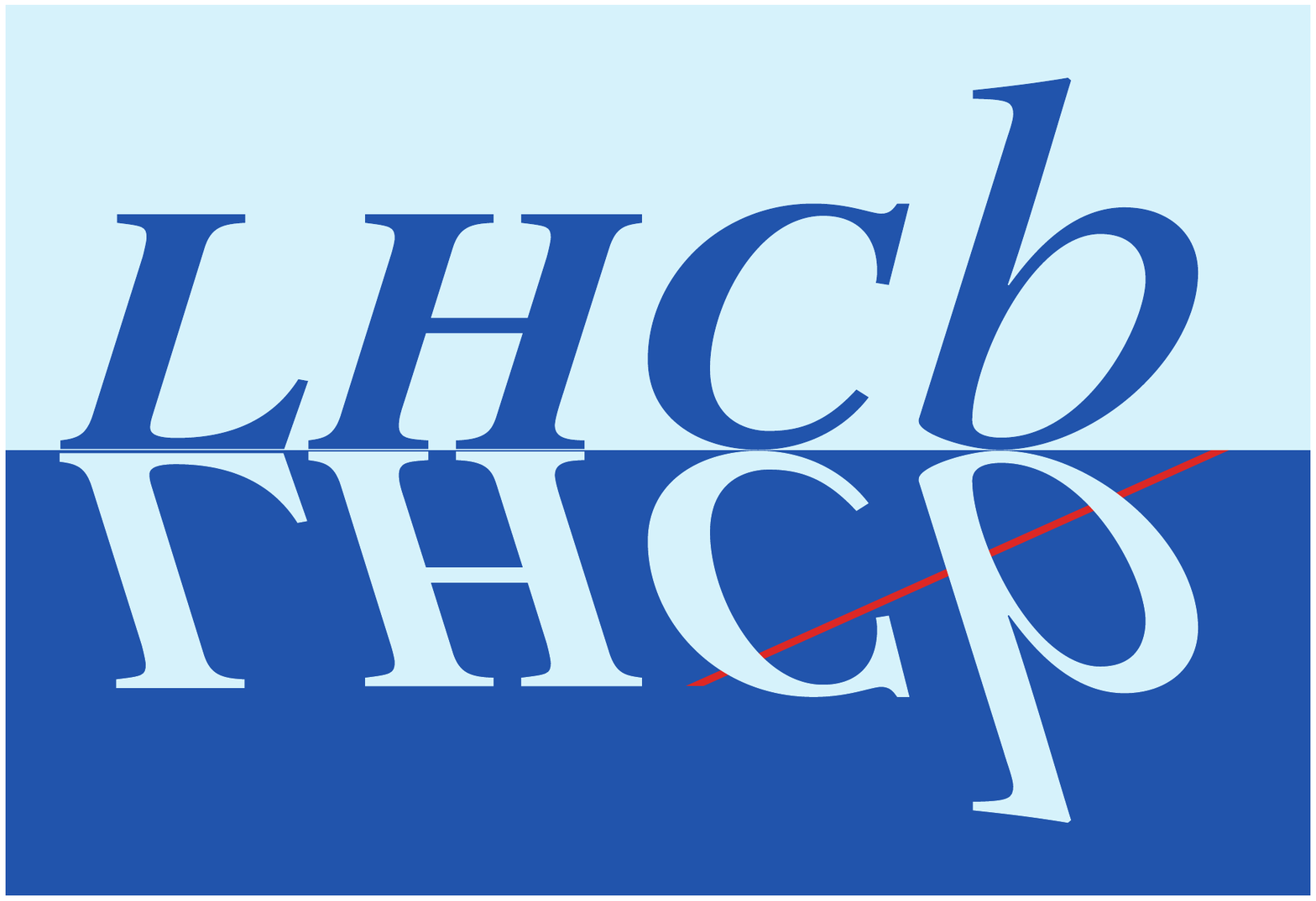}} & &}%
{\vspace*{-1.2cm}\mbox{\!\!\!\includegraphics[width=.12\textwidth]{lhcb-logo.eps}} & &}%
\\
 & & CERN-PH-EP-2014-140 \\  
 & & LHCb-PAPER-2014-024 \\  
 & & 24 June 2014
\end{tabular*}

\vspace*{4.0cm}

{\bf\boldmath\huge
\begin{center}
 Test of lepton universality using $B^{+}\rightarrow K^{+}\ell^{+}\ell^{-}$ decays 
\end{center}
}

\vspace*{2.0cm}

\begin{center}
The LHCb collaboration\footnote{Authors are listed on the following pages.}
\end{center}

\vspace{\fill}

\begin{abstract}
  \noindent
A measurement of the ratio of the branching fractions of the 
$B^{+}\rightarrow K^{+}\mu^{+}\mu^{-}$ and  $B^{+}\rightarrow K^{+}e^{+}e^{-}$ decays is presented using proton-proton collision data, corresponding to an integrated luminosity of 3.0\ensuremath{\mbox{\,fb}^{-1}}\xspace, 
recorded with the LHCb experiment at center-of-mass energies of 7 and 8 $\mathrm{\,Te\kern -0.1em V}$. 
The value of the ratio of branching fractions for the dilepton invariant mass squared range $1<q^{2}<6\mathrm{\,Ge\kern -0.1em V^2\!/}c^4$ is measured to be
$0.745^{+0.090}_{-0.074}\mathrm{\,(stat)}\,\pm0.036\mathrm{\,(syst)}$. 
This value is the most precise measurement of the ratio of branching fractions to date and is compatible with the SM prediction within $2.6$ standard deviations.
\end{abstract}

\vspace*{2.0cm}

\begin{center}
  Submitted to Phys. Rev. Lett. 
\end{center}

\vspace{\fill}

{\footnotesize 
\centerline{\copyright~CERN on behalf of the \lhcb collaboration, license \href{http://creativecommons.org/licenses/by/3.0/}{CC-BY-3.0}.}}
\vspace*{2mm}

\end{titlepage}


\newpage
\setcounter{page}{2}
\mbox{~}
\newpage

\centerline{\large\bf LHCb collaboration}
\begin{flushleft}
\small
R.~Aaij$^{41}$, 
B.~Adeva$^{37}$, 
M.~Adinolfi$^{46}$, 
A.~Affolder$^{52}$, 
Z.~Ajaltouni$^{5}$, 
S.~Akar$^{6}$, 
J.~Albrecht$^{9}$, 
F.~Alessio$^{38}$, 
M.~Alexander$^{51}$, 
S.~Ali$^{41}$
% ...

{\footnotesize \it
$ ^{1}$Centro Brasileiro de Pesquisas F\'{i}sicas (CBPF), Rio de Janeiro, Brazil\\
$ ^{2}$Universidade Federal do Rio de Janeiro (UFRJ), Rio de Janeiro, Brazil\\
% ...
$ ^{5}$Clermont Universit\'{e}, Universit\'{e} Blaise Pascal, CNRS/IN2P3, LPC, Clermont-Ferrand, France\\
$ ^{6}$CPPM, Aix-Marseille Universit\'{e}, CNRS/IN2P3, Marseille, France\\
% ...
$ ^{9}$Fakult\"{a}t Physik, Technische Universit\"{a}t Dortmund, Dortmund, Germany\\
% ...
$ ^{37}$Universidad de Santiago de Compostela, Santiago de Compostela, Spain\\
$ ^{38}$European Organization for Nuclear Research (CERN), Geneva, Switzerland\\
% ...
$ ^{41}$Nikhef National Institute for Subatomic Physics, Amsterdam, The Netherlands\\
% ...
$ ^{46}$H.H. Wills Physics Laboratory, University of Bristol, Bristol, United Kingdom\\
% ...
$ ^{51}$School of Physics and Astronomy, University of Glasgow, Glasgow, United Kingdom\\
$ ^{52}$Oliver Lodge Laboratory, University of Liverpool, Liverpool, United Kingdom\\
% ...
}
\end{flushleft}

\cleardoublepage


\renewcommand{\thefootnote}{\arabic{footnote}}
\setcounter{footnote}{0}


\pagestyle{plain} 
\setcounter{page}{1}
\pagenumbering{arabic}

\clearpage

The decay \BuKll, where \lepton represents either a muon or an electron, is a $\bquark\to\squark$ flavor-changing neutral current process.
Such processes are highly suppressed in the Standard Model (SM)  as they proceed through amplitudes involving electroweak loop (penguin and box) diagrams.
This makes the branching fraction of \BuKll\footnote{The inclusion of charge conjugate processes is implied throughout this Letter.} decays highly sensitive to the presence of virtual particles that are predicted to exist in extensions of the SM~\cite{Bobeth:2007dw}.
The decay rate of \BuKmm has been measured by \lhcb to a precision of 5\%~\cite{LHCb-PAPER-2012-024} and, 
 although the current theoretical uncertainties in the branching fraction are $\mathcal{O}(30\%)$~\cite{Bobeth:2011nj}, 
 these largely cancel in asymmetries or ratios of \BuKll observables~\cite{LHCb-PAPER-2013-043,LHCb-PAPER-2012-024,LHCb-PAPER-2012-011}.

Owing to the equality of the electroweak couplings of electrons and muons in the SM, known as lepton universality, the ratio of the branching fractions of \BuKmm to \BuKee decays~\cite{PhysRevD.69.074020} 
 is predicted to be unity within an uncertainty of $\mathcal{O}(10^{-3})$ in the SM~\cite{Bobeth:2007dw,Bouchard:2013mia}. 
The ratio of the branching fractions is particularly sensitive to extensions of the SM that introduce new scalar or pseudoscalar interactions~\cite{Bobeth:2007dw}. 
Models that contain a $Z'$ boson have recently been proposed to explain measurements of the angular distribution and branching fractions of \BdToKstmm and \BuKmm decays~\cite{Gauld,*Buras,*Altmannshofer}.
These types of models can also affect the relative branching fractions of \BuKll decays if the $Z'$ boson does not couple equally to electrons and muons.

Previous measurements of the ratio of branching fractions from \epem colliders operating at the $\PUpsilon(4S)$ resonance have measured values consistent with unity with a precision of 20--50\%~\cite{Lees:2012tva,*Wei:2009zv}. 
This Letter presents the most precise measurement of the ratio of branching fractions and the corresponding branching fraction \BR(\BuKee) to date.
The data used for these measurements are recorded in proton-proton (\proton{}\proton) collisions and correspond to 3.0\invfb of integrated luminosity, collected by the \lhcb experiment at center-of-mass energies of 7 and 8\tev.

The value of \RK within a given range of the dilepton mass squared from $q^{2}_{\rm min}$ to $q^{2}_{\rm max}$ is given by 
\begin{equation}
\label{eq:rk}
\RK  =  \dfrac{\int_{\qsq_{\rm min}}^{\qsq_{\rm max}} \dfrac{\deriv\Gamma[\BuKmm]}{\dqsq} \dqsq}{\int_{\qsq_{\rm min}}^{\qsq_{\rm max}} \dfrac{\deriv\Gamma[\BuKee]}{\deriv\qsq} \dqsq} ~,
\end{equation}
\noindent where $\Gamma$ is the \qsq-dependent partial width of the decay. 
We report a measurement of \RK for $1<\qsq<6\gevgevcccc$.
This range is both experimentally and theoretically attractive as it excludes the \BuJpsiKll resonant region, and precise theoretical predictions are possible. 
The high \qsq region, above the \psitwos resonance, is affected by broad charmonium resonances that decay to lepton pairs~\cite{LHCb-PAPER-2013-039}.

The value of \RK is determined using the ratio of the relative branching fractions of the decays \BuKll and \BuJpsiKll, with $\ell=e$ and $\mu$, respectively.
This takes advantage of the large \BuJpsiK branching fraction to cancel potential sources of systematic uncertainty between the \BuKll and \BuJpsiKll decays as the efficiencies are correlated and the branching fraction to \BuJpsiK is known precisely~\cite{PDG2012}. 
 This is achieved by using the same selection for \BuKll and \BuJpsiKll decays for each leptonic final state and by assuming lepton universality in the branching fractions of \jpsi mesons to the \mumu and \epem final states~\cite{PDG2012}.  
In terms of measured quantities \RK is written as
\begin{equation}
\label{eq:rk2}
\RK =
\left(\frac{\mathcal{N}_{\Kmm}}{\mathcal{N}_{\Kee}}\right)
\left(\frac{\mathcal{N}_{\JpsiKee}}{\mathcal{N}_{\JpsiKmm}}\right)
\left(\frac{\epsilon_{\Kee}}{\epsilon_{\Kmm}}\right)
\left( \frac{\epsilon_{\JpsiKmm}}{\epsilon_{\JpsiKee}}\right) ~,
\end{equation}
\noindent where $\mathcal{N}_{X}$ is the observed yield in final state $X$, and $\epsilon_{X}$ is the efficiency to trigger, reconstruct and select that final state.
Throughout this paper the number of \Kmm and \Kee candidates always refers to the restricted \qsq range, $1<\qsq<6\gevgevcccc$.

The \lhcb detector is a single-arm forward
spectrometer covering the \mbox{pseudorapidity} range $2<\eta <5$ and is described in detail in Ref.~\cite{Alves:2008zz}.
The simulated events used in this analysis are produced using the software described in Refs.~\cite{Sjostrand:2007gs,*LHCb-PROC-2010-056,*Lange:2001uf,*Golonka:2005pn,*Allison:2006ve,*Agostinelli:2002hh,*LHCb-PROC-2011-006}.

Candidate \BuKll events are first required to pass the hardware trigger that selects either muons with a high transverse momentum (\pt) or large energy deposits in the electromagnetic or hadronic calorimeters, which are a signature of high-\pt electrons or hadrons.
Events with muons in the final state are required to be triggered by one or both muons in the hardware trigger.
Events with electrons in the final state are required to be triggered by either one of the electrons, the kaon from the \Bp decay, or by other particles in the event. 
In the subsequent software trigger, at least one of the final-state particles is required to both have $\pt>800\mevc$ and not to originate from any of the primary $pp$ interaction vertices~(PVs) in the event. 
Finally, the tracks of the final-state particles are required to form a vertex that is significantly displaced from the PVs.
A multivariate algorithm~\cite{BBDT} is used for the identification of secondary vertices consistent with the decay of a \bquark hadron.

A \Kll candidate is formed from a pair of well-reconstructed oppositely charged particles identified as either electrons or muons, combined with another track that is identified as a charged kaon.
Each particle is required to have $\pt > 800\mevc$ and be inconsistent with coming from any PV.
The two leptons are required to originate from a common vertex, which is significantly displaced from all of the PVs in the event.
The \Kll candidate is required to have a good vertex fit, and the \Kll candidate is required to point to the best PV, defined by the lowest impact parameter (IP).

Muons are initially identified by tracks that penetrate the calorimeters and the iron filters in the muon stations~\cite{LHCB-DP-2013-001}. 
Further muon identification is performed with a multivariate classifier that uses information from the tracking system, the muon chambers, the ring-imaging Cherenkov (RICH) detectors and the calorimeters to provide separation of muons from pions and kaons. 
Electron identification is provided by matching tracks to an electromagnetic calorimeter (ECAL) cluster, combined with information from the RICH detectors, to build an overall likelihood for separating electrons from pions and kaons.

Bremsstrahlung from the electrons can significantly affect the measured electron momentum and the reconstructed \Bp candidate mass. 
To improve the accuracy of the electron momentum reconstruction, a correction for the measured momenta of photons associated to the electron is applied.
If an electron radiates a photon downstream of the dipole magnet, the photon enters the same ECAL cells as the electron itself and the original energy of the electron is measured by the ECAL.
However, if an electron  radiates a photon upstream of the dipole magnet, the energy of the photon will not be deposited in the same ECAL cells as the electron. 
After correction, the ratio of electron energy to the momentum measured by the ECAL is expected to be consistent with unity; the ratio is used in the electron identification likelihood.
Since there is little material within the magnet for particle interactions to cause additional neutral particles, the ECAL cells without an associated track are used to look for bremsstrahlung photons. 
A search is made for photons with transverse energy greater than 75\mev within a region of the ECAL defined by the extrapolation of the electron track upstream of the magnet.

The separation of the signal from combinatorial background uses a multivariate algorithm based on boosted decision trees~(BDT)~\cite{Breiman,*Roe}.  
 Independent BDTs are trained to separate the dielectron and dimuon signal decays from combinatorial backgrounds. 
The BDTs are trained using \BuJpsiKmm and \BuJpsiKee candidates in data to represent the signal, and candidates with ${\Kll}$ masses $m(\Kll) >  5700\mevcc$ as the background sample.
 The latter sample is not used in the subsequent analysis. 
 The variables used as input to the BDTs are: the transverse momentum of the \Bp candidate and of the final state particles; the \Bp decay time; the vertex fit quality; the IP of the \Bp candidate; the angle between the \Bp candidate momentum vector and direction between the best PV and the decay vertex; the IP of the final-state particles to the best PV and the track fit quality. 
 The most discriminating variable is the vertex quality for the \Bp and the angle between the \Bp candidate and the best PV. 
 The selections are optimized for the significance of the signal yield for each \BuKll decay and accept 60--70\% of the signal, depending on the decay channel, whilst rejecting over 95\% of the combinatorial background.  
The efficiency of the BDT response is uniform across the \qsq region of interest and in the \jpsi region, ensuring that the selection is not  significantly biased by the use of the \BuJpsiKll data. 

After applying the selection criteria, exclusive backgrounds from \bquark-hadron decays are dominated by three sources. 
The first is mis-reconstructed \BuJpsiKll and \BuPsiKll decays where the kaon is mistakenly identified as a lepton and the lepton (of the same electric charge) as a kaon. 
Such events are excluded using different criteria  for the muon and the electron modes owing to the lower momentum resolution in the latter case.  
The \BuKmm candidates are kept if the kaon passes through the acceptance of the muon detectors and is not identified  
 as a muon, or if the mass of the kaon candidate (in the muon mass hypothesis) and the oppositely charged muon candidate pair is distinct from the \jpsi or the \psitwos resonances.  
The \BuKee candidates are kept if the kaon has a low probability of being an electron according to the information from the electromagnetic and hadronic calorimeters and the RICH system. 
The second source of background is from semileptonic decays such as
\decay{\Bu}{\Dzb(\to\Kp\pim)\ellp\neul}, or 
\decay{\Bu}{\Dzb\pip}, with  \decay{\Dzb}{\Kp\ellm\neulb} or $\pip\ellm\neulb$,
 which can be selected as 
 signal decays if at least one of the hadrons is mistakenly identified as a lepton. 
All of these decays are vetoed by requiring that the mass of the $\Kp\ellm$ pair, where the lepton is assigned the pion mass, is greater than 1885\mevcc. 
These vetoes result in a negligible loss of signal as measured in simulation.
The third source of background is partially reconstructed \bquark-hadron decays that are reconstructed with masses smaller than the measured \Bp mass.  
In the muon decay modes, this background is excluded by the choice of $m(\Kp\mumu)$ mass interval, while in the electron modes this background is described in the mass fit model.  
Fully hadronic \bquark-meson decays, such as \decay{\Bp}{\Kp\pip\pim}, are reduced to $\mathcal{O}(0.1\%)$ of the \BuKmm and \BuKee signals by the electron and muon identification requirements, respectively, and are neglected in the analysis.  

\begin{figure}[tbp]
\centering
\includegraphics[width=0.495\textwidth]{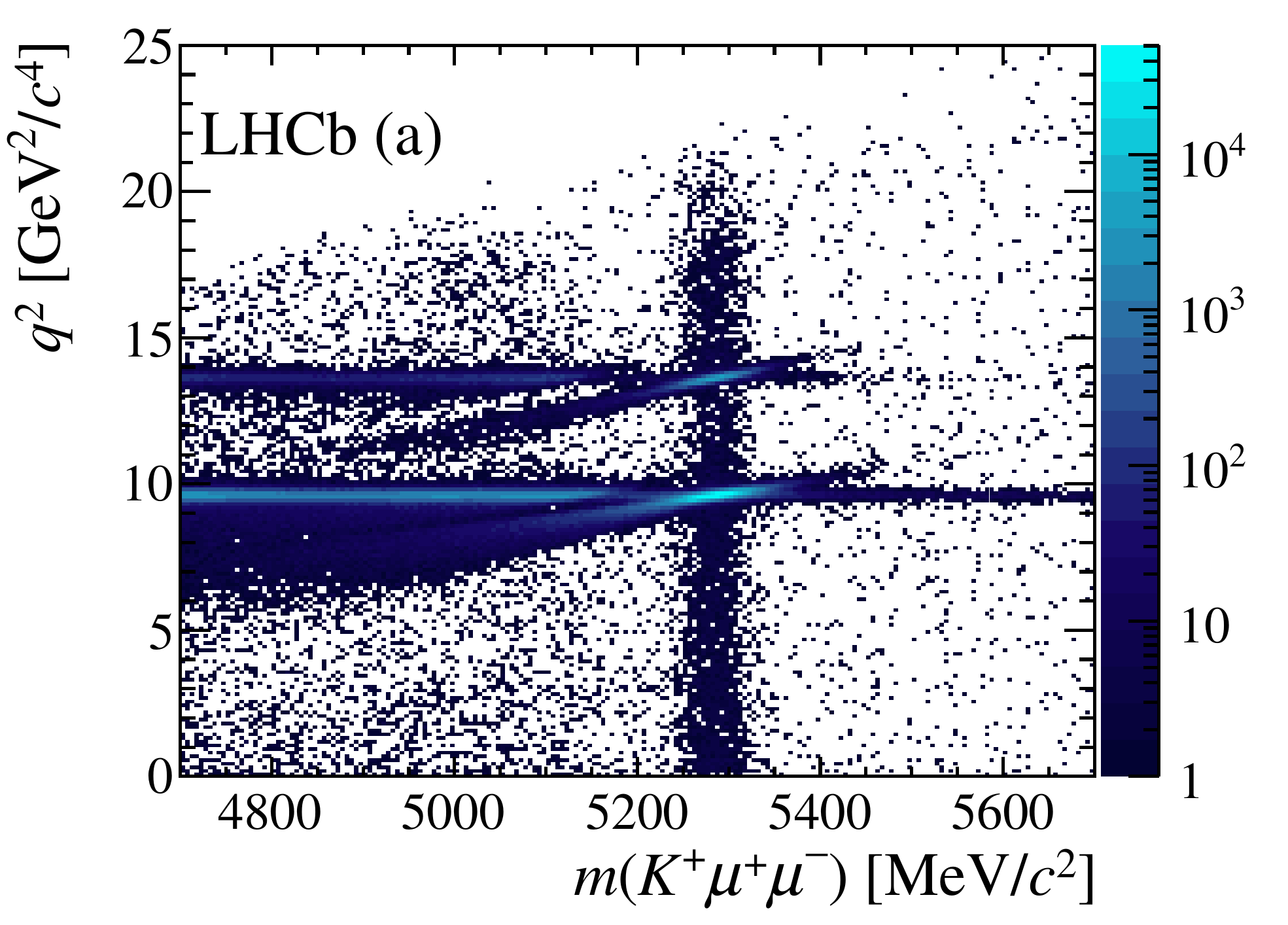}
\includegraphics[width=0.495\textwidth]{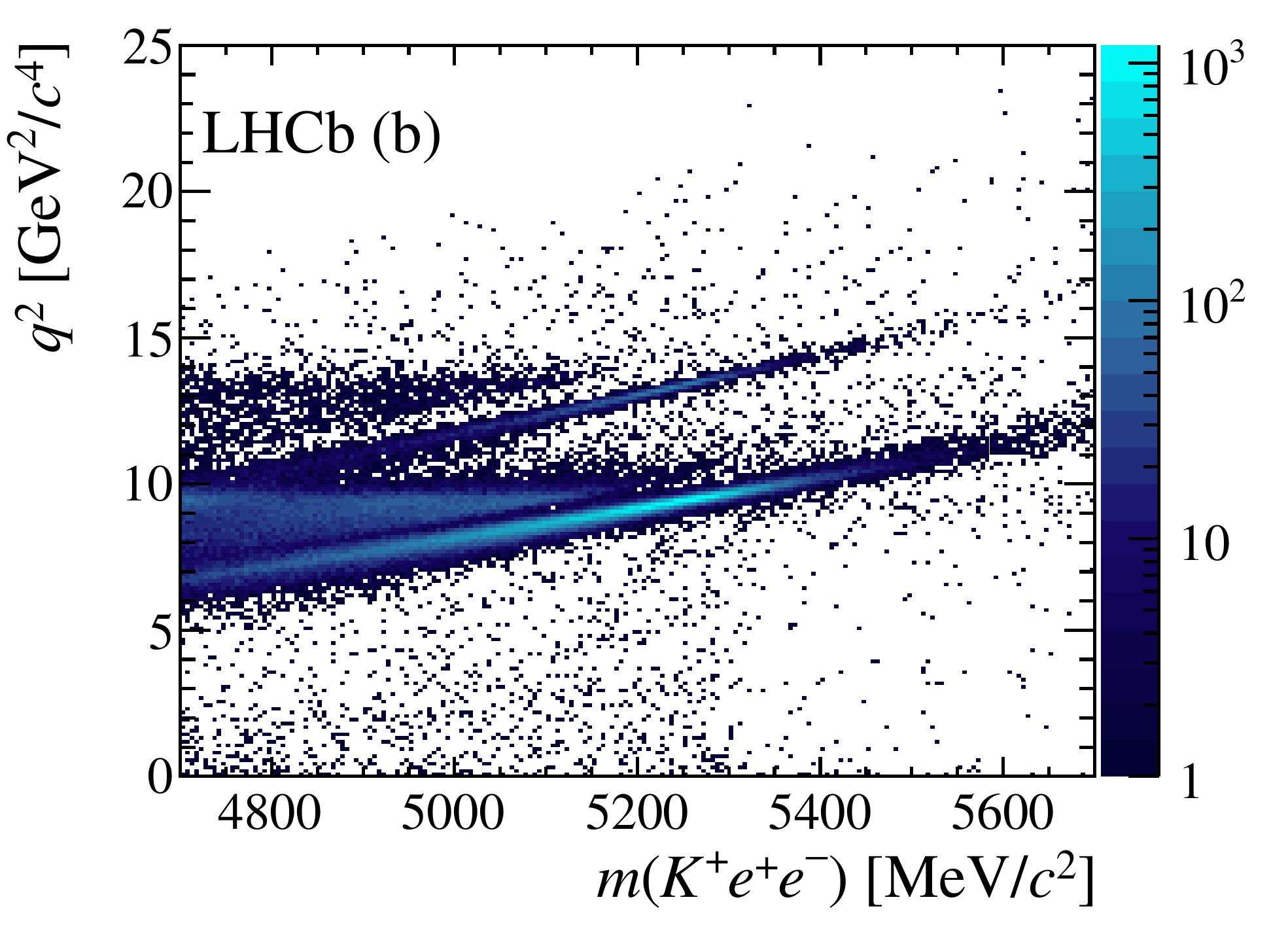}
\caption{Dilepton invariant mass squared, \qsq, as a function of the \Kll invariant mass, $m(\Kll)$, for selected (a) \BuKmm and (b) \BuKee candidates. 
The radiative tail of the \jpsi and \psitwos mesons is most pronounced in the electron mode due to the larger bremsstrahlung and because the energy resolution of the ECAL is lower compared to the momentum resolution of the tracking system. 
\label{fig:twodplots}}
\end{figure}

The reconstructed \Bp mass and dilepton mass of the candidates passing the selection criteria are shown in Fig.~\ref{fig:twodplots}. 
It is possible to see the pronounced peaks of the \jpsi and \psitwos decays along with their radiative tail as a diagonal band. 
Partially reconstructed decays can be seen to lower \Kll masses and the distribution of random combinatorial background at high \Kll masses.
Only candidates with $5175< m({\Kmm}) <5700\mevcc$ or $4880< m(\Kee) <5700\mevcc$ are considered. 
The dilepton mass squared is also restricted to $1<\qsq<6\gevgevcccc$,  $8.68<\qsq<10.09\gevgevcccc$ and  $6<\qsq<10.09\gevgevcccc$ when selecting \BuKll,  \BuJpsiKmm and \BuJpsiKee candidates, respectively.

The event yields for the \BuKll and the \BuJpsiKll modes are determined using unbinned extended maximum likelihood fits to the \Kll mass distributions.
The model is composed of a signal shape, a combinatorial background shape and, for the electron modes, a contribution from partially reconstructed \bquark-hadron decays. 

The signal mass model for the muon modes consists of the sum of two Crystal Ball functions~\cite{Skwarnicki:1986xj} with tails above and below the mass peak.
This empirical function describes the core of the mass distribution and additional effects from the experimental resolution and the radiative tail.
The mean, width and radiative tail parameters for the signal model are obtained from a fit to the \BuJpsiKmm sample and propagated to the fit for the \BuKmm decays.
The validity of this approach is verified using simulation.
The combinatorial background is described by an exponential function.
There are $667\,046\pm882$ \BuJpsiKmm and $1226\pm41$ \BuKmm signal decays, where the uncertainties are statistical.

The mass distribution of the electron modes depends strongly on the number of bremsstrahlung photons that are associated with the electrons, and therefore a more involved parametrization is required. 
The mass distribution also depends on the \pt of the electrons and on the occupancy of the event. 
This shape dependence is studied using a selection of \BuJpsiKee events in the data.
The data are split into three independent samples according to which particle in the event has fired the hardware trigger; a similar strategy was applied in Ref.~\cite{LHCb-PAPER-2013-005}.  
These categories are mutually exclusive and consist of events selected either by one of the two electrons, by the \Kp meson, or by other particles. 
Events that are triggered by one of the electrons in the hardware trigger typically have larger electron momentum and \pt than events triggered by the \Kp meson or other particles in the event. 
Within each of these trigger categories, independent shapes are used depending on the number of neutral clusters that are added to the dielectron candidate to correct for the effects of bremsstrahlung:
 one for candidates where no clusters are added to either electron; one for candidates where a cluster is added to one of the electrons; and one for candidates where clusters are added to both electrons.  
The fractions of candidates in each of these categories are 37\%, 48\% and 15\%, respectively, for both \BuJpsiKee and \BuKee candidates. 
The relative proportion of the three categories for the number of additional clusters is described well by the simulation. 
 Candidates with no added clusters have a large radiative tail to smaller $m(\Kee)$ values. 
 Candidates with one or more added clusters have a reduced radiative tail, but have larger tails above the expected \Bp mass due to the event occupancy or the resolution of the ECAL. 

The parametrization of the \BuKee mass distribution in each of the three trigger categories is described by a sum of three Crystal Ball functions, each of which has independent values for the peak, width and radiative tail, representing the different number of clusters that are added.
 The parameters for each of the Crystal Ball functions are found by fitting the $m(\Kee)$ distribution of the \BuJpsiKee candidates. 
 A high-purity sample of \BuJpsiKee candidates is achieved by constraining the mass of the \epem pair to the known \jpsi mass. 
A requirement that $m(\jpsi\Kp)$ is greater than $5175\mevcc$ removes partially reconstructed signal candidates, leaving a prominent signal peak with negligible contribution from combinatorial backgrounds without biasing the mass shape. 

 The mass distribution of the partially reconstructed backgrounds is determined using simulated \decay{H_\bquark}{\jpsi(\to \epem)X} decays that satisfy the selection criteria, where $H_\bquark$ is a \Bu, \Bd, \Bs or \Lb hadron.
The relative branching fraction of \decay{H_\bquark}{\jpsi(\to \epem)X} to \decay{H_\bquark}{\epem X} decays is assumed to be the same as that of \BuJpsiKee and \BuKee decays, 
and is consistent with the observed ratios of \BuJpsiKmm to \BuKmm decays and \BdJpsiKstmm to \BdKstmm decays~\cite{PDG2012}.

The ratio of partially reconstructed background to signal for the decay \BuKee is determined by the ratio measured in \BuJpsiKee data for each trigger category, after correcting for two factors.
Firstly, the partially reconstructed backgrounds for the \BuJpsiK data may include a contribution from  cascade decays of higher \ccbar resonances, \eg, \decay{\Bp}{\psitwos (\to \jpsi \pip\pim)\Kp} or \decay{\Bp}{\chic (\to \jpsi \gamma)\Kp} decays.
These decays contribute to the \BuJpsiK background but not to the partially reconstructed backgrounds for the \BuKee data. 
The level of contamination is estimated using simulated inclusive \decay{H_\bquark}{\jpsi(\to\epem)X} decays and found to be $(16\pm1)\%$.
Secondly, the dominant contribution to the \BuKee background is from partially reconstructed \BdKstee decays. The relative proportion of \BdKstmm to \BuKmm 
 decays is known to be 10\% higher than the relative proportion of \BdJpsiKst to \BuJpsiK decays~\cite{PDG2012}. 
The fraction of partially reconstructed background  to signal  is adjusted accordingly.
The partially reconstructed backgrounds account for 16--20\% of the signal yields depending on the trigger category.

\begin{figure}[t]
\centering
\includegraphics[width=0.32\textwidth]{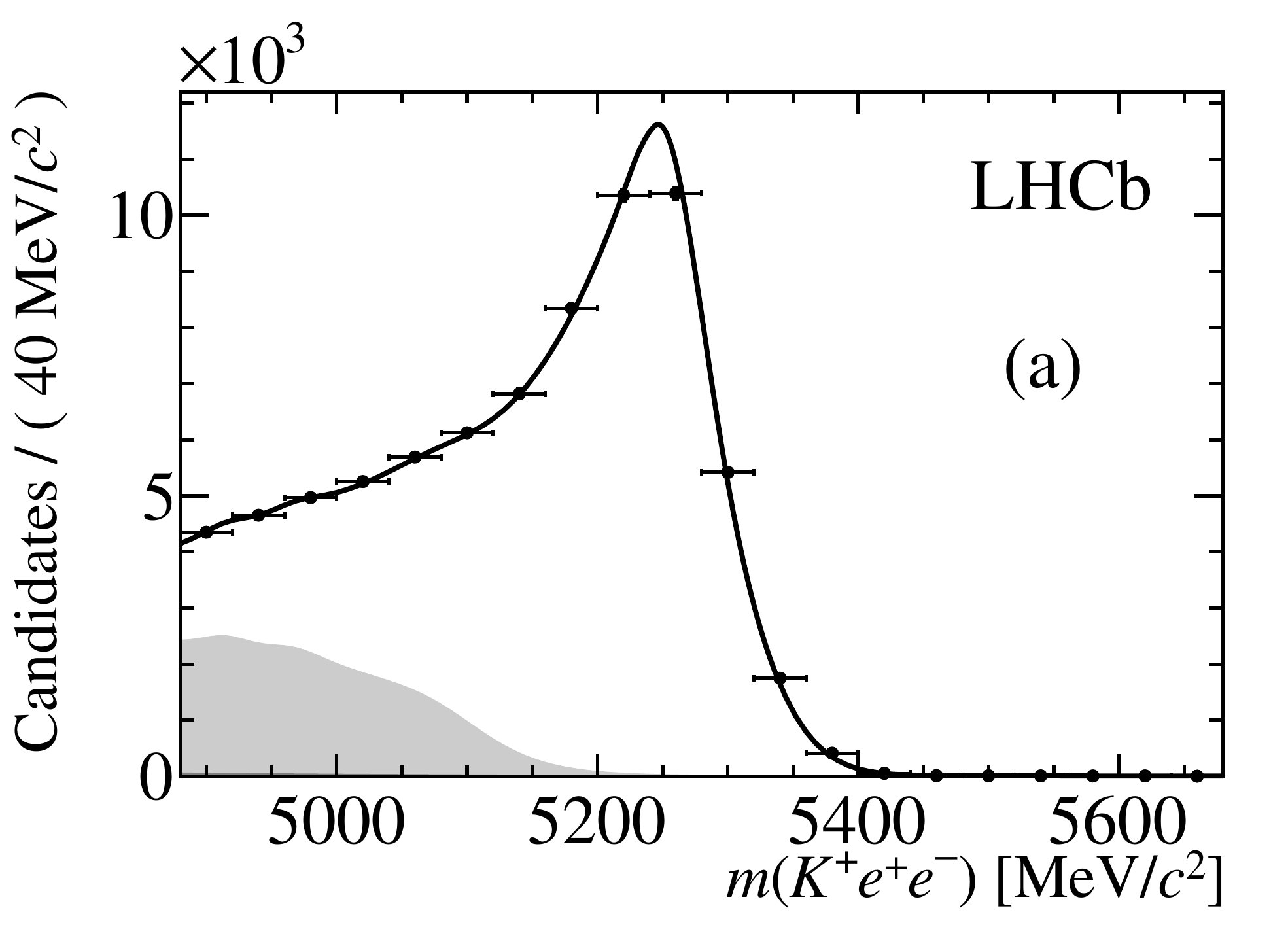}
\includegraphics[width=0.32\textwidth]{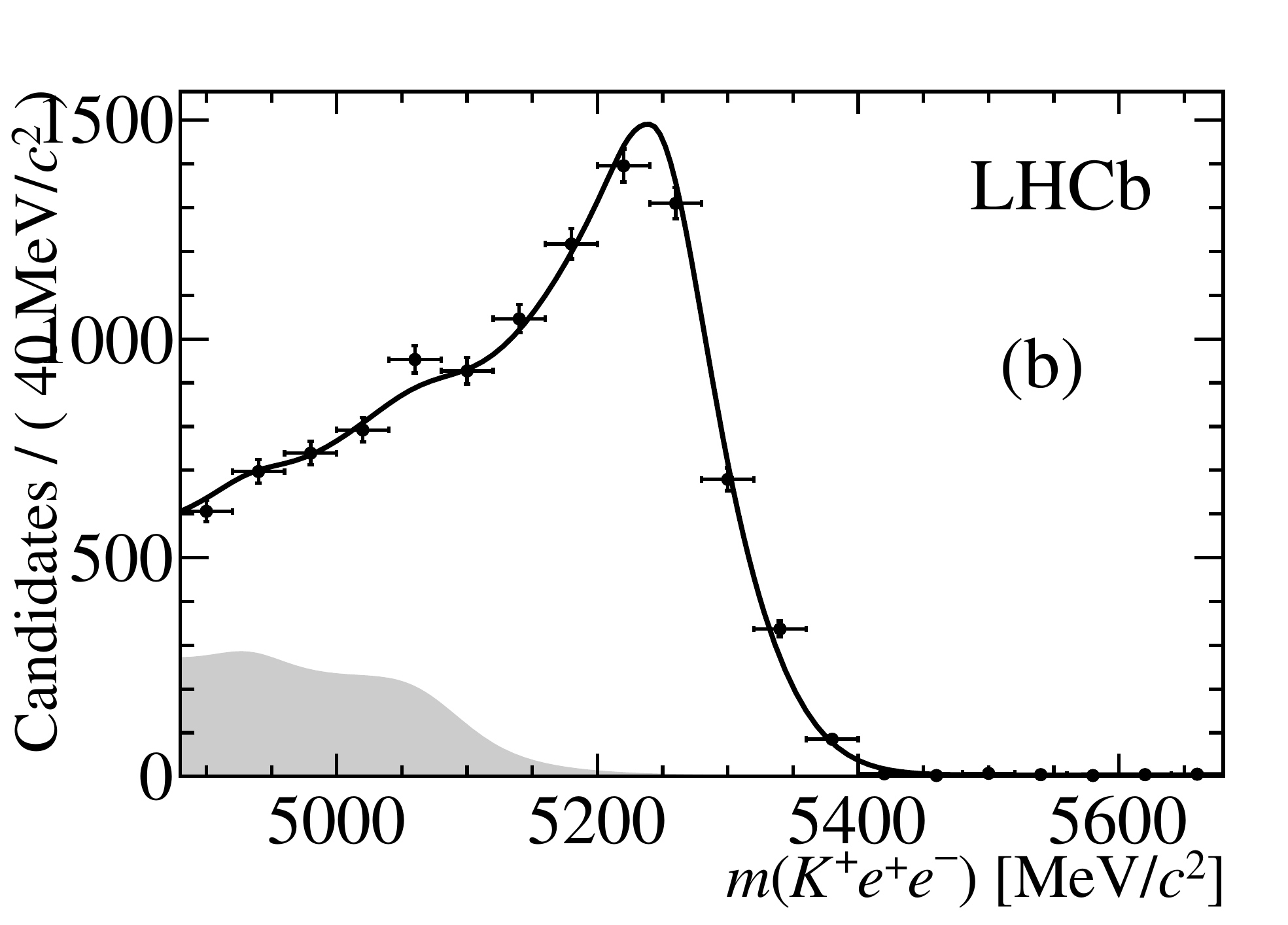} 
\includegraphics[width=0.32\textwidth]{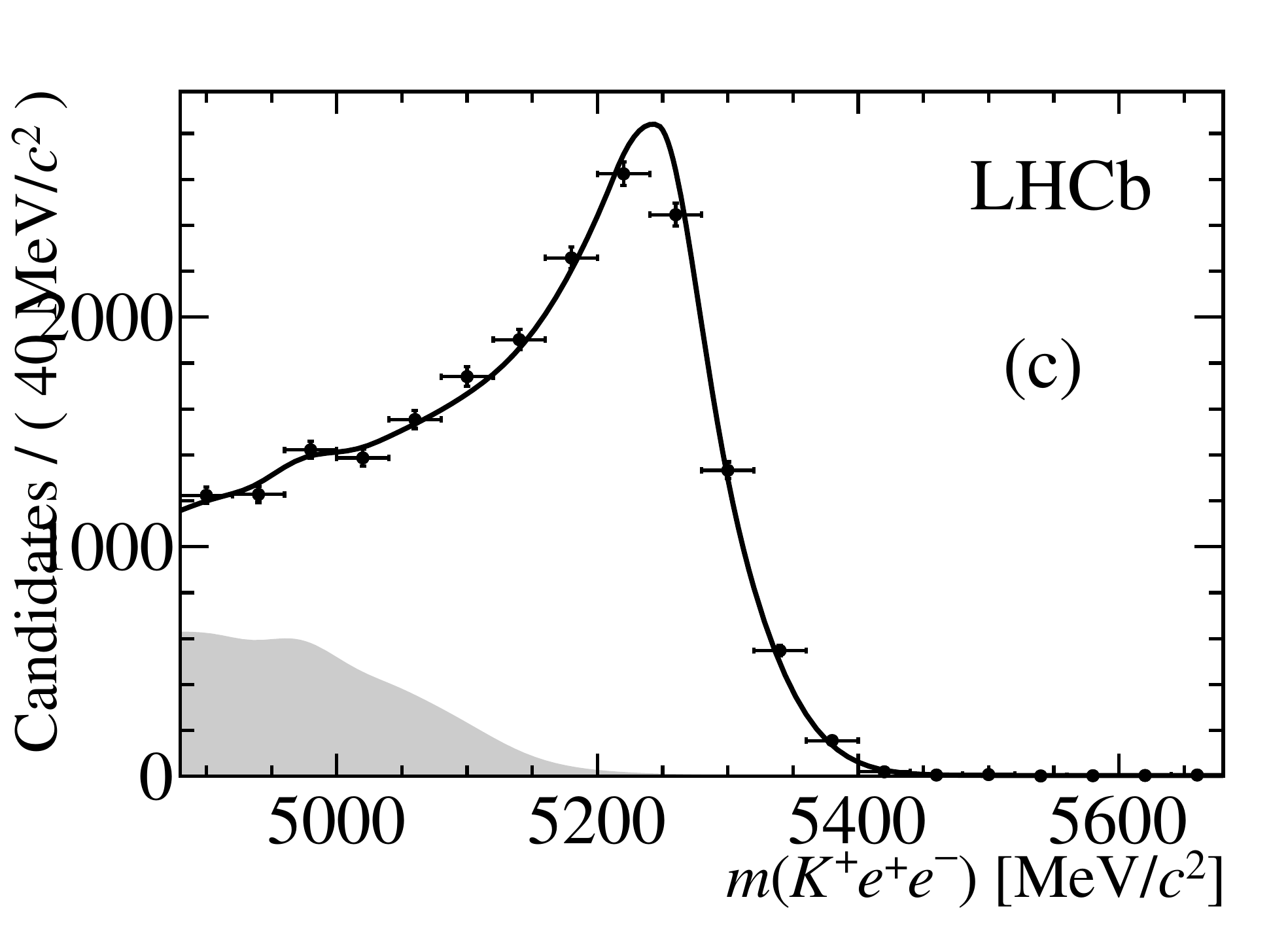}
\includegraphics[width=0.32\textwidth]{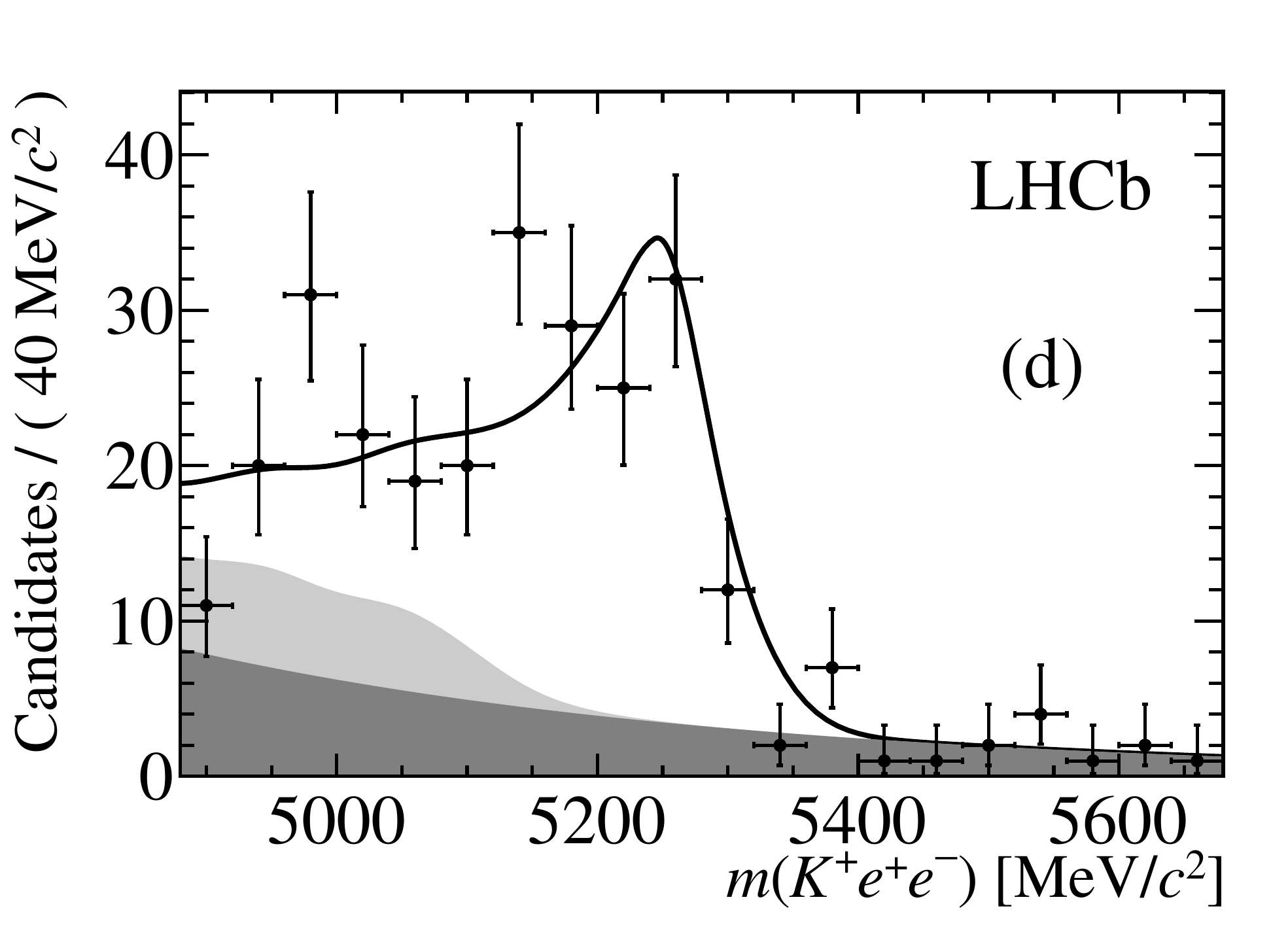}
\includegraphics[width=0.32\textwidth]{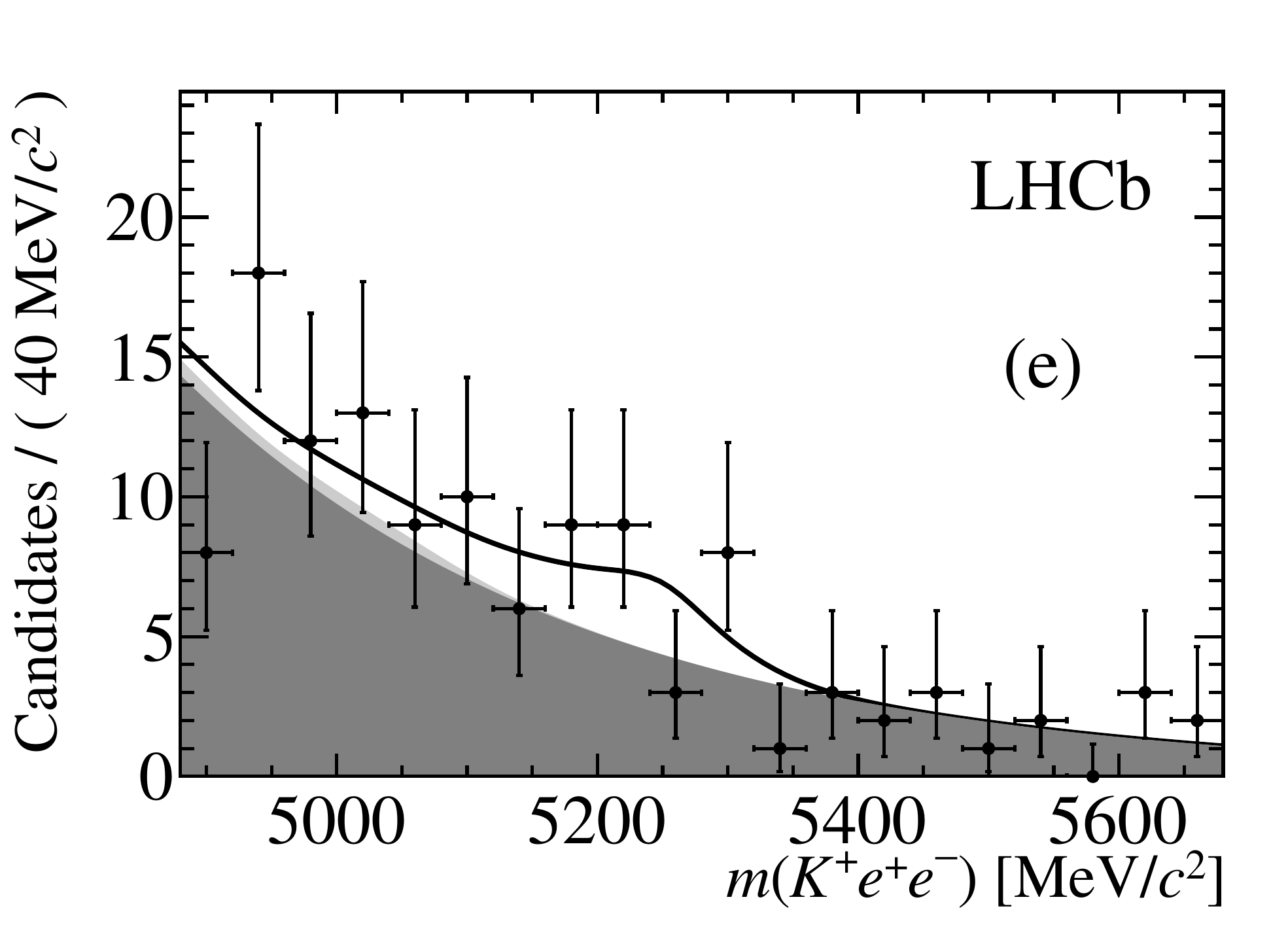}
\includegraphics[width=0.32\textwidth]{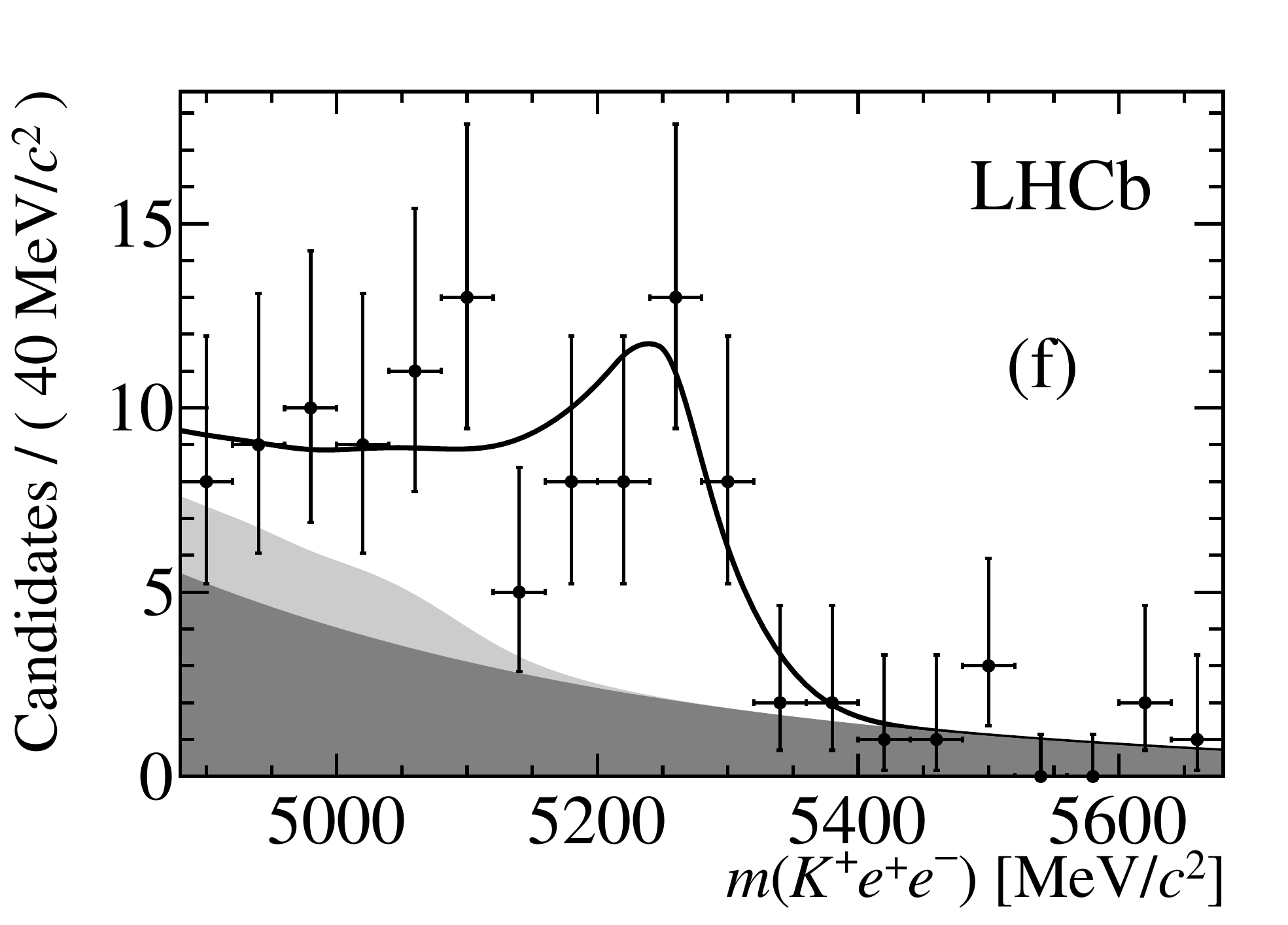}
\caption{Mass distributions with fit projections overlaid of selected \BuJpsiKee candidates triggered in the hardware trigger by (a) one of the two electrons, (b) by the \Kp and (c) by other particles in the event. 
 Mass distributions with fit projections overlaid of selected \BuKee candidates in the same categories, triggered by (d) one of the two electrons, (e) the \Kp and (f) by other particles in the event. 
The total fit model is shown in black, the combinatorial background component is indicated by the dark shaded region and the background from partially reconstructed \bquark-hadron decays by the light shaded region. \label{fig:electrons} 
}
\end{figure}

The results of the fits for the \BuJpsiKee and \BuKee channels are shown in Fig.~\ref{fig:electrons}. In total there are $172^{+20}_{-19}$ ($62~324\pm318$)  \BuKee (\BuJpsiKee) decays triggered by the electron trigger,
 $20^{+16}_{-14}$ ($9~337\pm124$) decays triggered by the hadron trigger and $62\pm13$ ($16~796\pm165$) decays that were triggered by other particles in the event. 

It is possible for \BuKee decays that emit bremsstrahlung to migrate out of the $1<\qsq<6\gevgevcccc$ range at the lower edge and in from the upper edge.
The effect of this bin migration on the yield is determined using \BuKee simulation and validated with \BuJpsiKee data.
The corresponding uncertainty due to the dependence of the branching fraction on non-SM contributions is estimated by independantly varying the \Bu\to\Kp form factors and by adjusting the Wilson coefficients \cite{Bobeth:2010wg}.
The overall yield of \BuKee is scaled by $(90.9\pm1.5)\%$ to account for this migration, where the uncertainty is mainly due to the model dependence.
The quality of the fits to the mass distribution of \Kll candidates is investigated and found to be acceptable. 

The systematic dependence of the signal yield on the signal model is considered negligible for the muon modes due to the excellent dimuon mass 
resolution at \lhcb \cite{LHCb-PAPER-2011-035,*LHCb-PAPER-2012-048}. 
The proportion of the partially reconstructed backgrounds is changed based on the measurements of the \BuJpsiXee contribution in Refs.~\cite{LHCb-PAPER-2012-010,LHCb-PAPER-2013-024} and contributes a systematic uncertainty of 1.6\% to the value of \RK.
The uncertainty in the signal model for the \BuKee mass distribution is assessed by incorporating a resolution effect 
 that takes into account the difference between the mass shape in simulated events for \BuJpsiKee and \BuKee decays
 and contributes a relative systematic uncertainty of 3\% to the value of \RK.

The efficiency to select \BuKmm, \BuKee, \BuJpsiKmm and \BuJpsiKee decays is the product of the efficiency to reconstruct the final state particles. 
This includes the geometric acceptance of the detector, the trigger and the selection efficiencies. 
Each of these efficiencies is determined from simulation and is corrected for known differences relative to data. 
 The use of the double ratio of decay modes ensures that most of the possible sources of systematic uncertainty cancel when determining \RK. 
 Residual effects from the trigger and the particle identification that do not cancel in the ratio arise due to different final-state particle kinematic distributions in the resonant and non resonant dilepton mass region.

The dependence of the particle identification on the kinematic distributions contributes a systematic uncertainty of 0.2\% to the value of \RK. 
The efficiency associated with the hardware trigger on \BuJpsiKee and \BuKee decays depends strongly on the kinematic properties of the final state particles and does not entirely cancel in the calculation of \RK, due to different electron and muon trigger thresholds.
 The efficiency associated with the hardware trigger is determined using simulation and is cross-checked using \BuJpsiKee and \BuJpsiKmm candidates in the data, 
by comparing candidates triggered by the kaon or leptons in the hardware trigger to candidates triggered by other particles in the event. 
The largest difference between data and simulation in the ratio of trigger efficiencies between the \BuKll and \BuJpsiKll decays is at the level of 3\%, which is assigned as a systematic uncertainty on \RK. 
The veto to remove misidentification of kaons as electrons contains a similar dependence on the chosen binning scheme and a systematic uncertainty of 0.6\% on \RK is assigned to account for this.

Overall, the efficiency to reconstruct, select and identify an electron is around 50\% lower than the efficiency for a muon. 
The total efficiency in the range $1 < \qsq < 6\gevgevcccc$ is also lower for \BuKll decays than the efficiency for the \BuJpsiKll decays, due to the softer lepton momenta in this \qsq range.

The ratio of efficiency-corrected yields of \BuKee to \BuJpsiKee is determined separately for each type of hardware trigger and then combined with the ratio of efficiency-corrected yields for the muon decays. 
\RK is measured to have a value of $0.72^{+0.09}_{-0.08}\stat\pm0.04\syst$, $1.84^{+1.15}_{-0.82}\stat\pm0.04\syst$ and $0.61^{+0.17}_{-0.07}\stat\pm0.04\syst $ for dielectron events triggered by electrons, the kaon or other particles in the event, respectively.
Sources of systematic uncertainty are assumed to be uncorrelated and are added in quadrature. 
Combining these three independent measurements of \RK and taking into account correlated uncertainties from the muon yields and efficiencies, gives
\begin{displaymath}
\RK  = 0.745^{+0.090}_{-0.074}\stat\,\pm0.036\syst.
\end{displaymath}
The dominant sources of systematic uncertainty are due to the parameterization of the \BuJpsiKee mass distribution and the estimate of the trigger efficiencies that both contribute 3\% to the value of \RK.

The branching fraction of \BuKee is determined in the region from $1<\qsq<6\gevgevcccc$ by taking the ratio of the branching fraction from \BuKee and \BuJpsiKee decays and multiplying  it by the measured value of \BR(\BuJpsiK) and \decay{\jpsi}{\epem}~\cite{PDG2012}. The value obtained is $\BR(\BuKee) = (1.56\,^{+0.19}_{-0.15}\stat\,^{+0.06}_{-0.04}\syst) \times10^{-7}~ $. This is the most precise measurement to date and is consistent with the SM expectation.

In summary, the ratio of branching fractions for \BuKmm and \BuKee decays, \RK, is measured in the dilepton invariant mass squared range from $1<\qsq<6\gevgevcccc$ with a total precision of 10\%.
A new measurement of the differential branching fraction of \BuKee is also reported.
The value of \RK is the most precise measurement of this quantity to date.  
It is compatible with the SM expectation to within $2.6$ standard deviations calculated using the ratio of the likelihoods between central value and the SM prediction.  
\bigskip

We express our gratitude to our colleagues in the CERN
accelerator departments for the excellent performance of the LHC. We
thank the technical and administrative staff at the LHCb
institutes. We acknowledge support from CERN and from the national
agencies: CAPES, CNPq, FAPERJ and FINEP (Brazil); NSFC (China);
CNRS/IN2P3 (France); BMBF, DFG, HGF and MPG (Germany); SFI (Ireland); INFN (Italy); 
FOM and NWO (The Netherlands); MNiSW and NCN (Poland); MEN/IFA (Romania); 
MinES and FANO (Russia); MinECo (Spain); SNSF and SER (Switzerland); 
NASU (Ukraine); STFC (United Kingdom); NSF (USA).
The Tier1 computing centres are supported by IN2P3 (France), KIT and BMBF 
(Germany), INFN (Italy), NWO and SURF (The Netherlands), PIC (Spain), GridPP 
(United Kingdom).
We are indebted to the communities behind the multiple open 
source software packages on which we depend. We are also thankful for the 
computing resources and the access to software R\&D tools provided by Yandex LLC (Russia).
Individual groups or members have received support from 
EPLANET, Marie Sk\l{}odowska-Curie Actions and ERC (European Union), 
Conseil g\'{e}n\'{e}ral de Haute-Savoie, Labex ENIGMASS and OCEVU, 
R\'{e}gion Auvergne (France), RFBR (Russia), XuntaGal and GENCAT (Spain), Royal Society and Royal
Commission for the Exhibition of 1851 (United Kingdom).
 
\addcontentsline{toc}{section}{References}
\setboolean{inbibliography}{true}

\begin{mcitethebibliography}{10}
\mciteSetBstSublistMode{n}
\mciteSetBstMaxWidthForm{subitem}{\alph{mcitesubitemcount})}
\mciteSetBstSublistLabelBeginEnd{\mcitemaxwidthsubitemform\space}
{\relax}{\relax}

\bibitem{Bobeth:2007dw}
C.~Bobeth, G.~Hiller, and G.~Piranishvili,
  \ifthenelse{\boolean{articletitles}}{{\it {Angular distributions of $\bar{B}
  \rightarrow \bar{K} \bar{l} l $ decays}},
  }{}\href{http://dx.doi.org/10.1088/1126-6708/2007/12/040}{JHEP {\bf 12}
  (2007) 040}, \href{http://arxiv.org/abs/0709.4174}{{\tt
  arXiv:0709.4174}}\relax
\mciteBstWouldAddEndPuncttrue
\mciteSetBstMidEndSepPunct{\mcitedefaultmidpunct}
{\mcitedefaultendpunct}{\mcitedefaultseppunct}\relax
\EndOfBibitem
\bibitem{LHCb-PAPER-2012-024}
LHCb collaboration, R.~Aaij {\em et~al.},
  \ifthenelse{\boolean{articletitles}}{{\it {Differential branching fraction
  and angular analysis of the $B^{+} \to K^{+}\mu^{+}\mu^{-}$ decay}},
  }{}\href{http://dx.doi.org/10.1007/JHEP02(2013)105}{JHEP {\bf 02} (2013)
  105}, \href{http://arxiv.org/abs/1209.4284}{{\tt arXiv:1209.4284}}\relax
\mciteBstWouldAddEndPuncttrue
\mciteSetBstMidEndSepPunct{\mcitedefaultmidpunct}
{\mcitedefaultendpunct}{\mcitedefaultseppunct}\relax
\EndOfBibitem
\bibitem{Bobeth:2011nj}
C.~Bobeth, G.~Hiller, D.~van Dyk, and C.~Wacker,
  \ifthenelse{\boolean{articletitles}}{{\it {The decay $B \rightarrow K l^+
  l^-$ at low hadronic recoil and model-independent $\Delta B = 1$
  constraints}}, }{}\href{http://dx.doi.org/10.1007/JHEP01(2012)107}{JHEP {\bf
  01} (2012) 107}, \href{http://arxiv.org/abs/1111.2558}{{\tt
  arXiv:1111.2558}}\relax
\mciteBstWouldAddEndPuncttrue
\mciteSetBstMidEndSepPunct{\mcitedefaultmidpunct}
{\mcitedefaultendpunct}{\mcitedefaultseppunct}\relax
\EndOfBibitem
\bibitem{LHCb-PAPER-2013-043}
LHCb collaboration, R.~Aaij {\em et~al.},
  \ifthenelse{\boolean{articletitles}}{{\it {Measurement of the $CP$ asymmetry
  in $B^+\to K^+\mu^+\mu^-$ decays}},
  }{}\href{http://dx.doi.org/10.1103/PhysRevLett.111.151801}{Phys.\ Rev.\
  Lett.\  {\bf 111} (2013) 151801}, \href{http://arxiv.org/abs/1308.1340}{{\tt
  arXiv:1308.1340}}\relax
\mciteBstWouldAddEndPuncttrue
\mciteSetBstMidEndSepPunct{\mcitedefaultmidpunct}
{\mcitedefaultendpunct}{\mcitedefaultseppunct}\relax
\EndOfBibitem
\bibitem{LHCb-PAPER-2012-011}
LHCb collaboration, R.~Aaij {\em et~al.},
  \ifthenelse{\boolean{articletitles}}{{\it {Measurement of the isospin
  asymmetry in $B \to K^{(*)}\mu^ + \mu^-$ decays}},
  }{}\href{http://dx.doi.org/10.1007/JHEP07(2012)133}{JHEP {\bf 07} (2012)
  133}, \href{http://arxiv.org/abs/1205.3422}{{\tt arXiv:1205.3422}}\relax
\mciteBstWouldAddEndPuncttrue
\mciteSetBstMidEndSepPunct{\mcitedefaultmidpunct}
{\mcitedefaultendpunct}{\mcitedefaultseppunct}\relax
\EndOfBibitem
\bibitem{PhysRevD.69.074020}
G.~Hiller and F.~Kr\"uger, \ifthenelse{\boolean{articletitles}}{{\it {More
  model-independent analysis of $b\rightarrow s$ processes}},
  }{}\href{http://dx.doi.org/10.1103/PhysRevD.69.074020}{Phys.\ Rev.\  {\bf
  D69} (2004) 074020}, \href{http://arxiv.org/abs/hep-ph/0310219}{{\tt
  arXiv:hep-ph/0310219}}\relax
\mciteBstWouldAddEndPuncttrue
\mciteSetBstMidEndSepPunct{\mcitedefaultmidpunct}
{\mcitedefaultendpunct}{\mcitedefaultseppunct}\relax
\EndOfBibitem
\bibitem{Bouchard:2013mia}
C.~Bouchard {\em et~al.}, \ifthenelse{\boolean{articletitles}}{{\it {Standard
  Model predictions for $B \rightarrow K l^+ l^-$ with form factors from
  lattice QCD}},
  }{}\href{http://dx.doi.org/10.1103/PhysRevLett.111.162002}{Phys.\ Rev.\
  Lett.\  {\bf 111} (2013) 162002}, \href{http://arxiv.org/abs/1306.0434}{{\tt
  arXiv:1306.0434}}\relax
\mciteBstWouldAddEndPuncttrue
\mciteSetBstMidEndSepPunct{\mcitedefaultmidpunct}
{\mcitedefaultendpunct}{\mcitedefaultseppunct}\relax
\EndOfBibitem
\bibitem{Gauld}
R.~Gauld, F.~Goertz, and U.~Haisch, \ifthenelse{\boolean{articletitles}}{{\it
  {An explicit $Z^{'}$-boson explanation of the $B \to K^* \mu^+ \mu^-$
  anomaly}}, }{}\href{http://dx.doi.org/10.1007/JHEP01(2014)069}{JHEP {\bf 01}
  (2014) 069}, \href{http://arxiv.org/abs/1310.1082}{{\tt
  arXiv:1310.1082}}\relax
\mciteBstWouldAddEndPuncttrue
\mciteSetBstMidEndSepPunct{\mcitedefaultmidpunct}
{\mcitedefaultendpunct}{\mcitedefaultseppunct}\relax
\EndOfBibitem
\bibitem{Buras}
A.~J. Buras, F.~De~Fazio, J.~Girrbach, and M.~V. Carlucci,
  \ifthenelse{\boolean{articletitles}}{{\it {The anatomy of quark flavour
  observables in 331 models in the flavour precision era}},
  }{}\href{http://dx.doi.org/10.1007/JHEP02(2013)023}{JHEP {\bf 02} (2013)
  023}, \href{http://arxiv.org/abs/1211.1237}{{\tt arXiv:1211.1237}}\relax
\mciteBstWouldAddEndPuncttrue
\mciteSetBstMidEndSepPunct{\mcitedefaultmidpunct}
{\mcitedefaultendpunct}{\mcitedefaultseppunct}\relax
\EndOfBibitem
\bibitem{Altmannshofer}
W.~Altmannshofer and D.~M. Straub, \ifthenelse{\boolean{articletitles}}{{\it
  {New physics in $B \to K^*\mu\mu$ ?}},
  }{}\href{http://dx.doi.org/10.1140/epjc/s10052-013-2646-9}{Eur.\ Phys.\ J.\
  {\bf C73} (2013) 2646}, \href{http://arxiv.org/abs/1308.1501}{{\tt
  arXiv:1308.1501}}\relax
\mciteBstWouldAddEndPuncttrue
\mciteSetBstMidEndSepPunct{\mcitedefaultmidpunct}
{\mcitedefaultendpunct}{\mcitedefaultseppunct}\relax
\EndOfBibitem
\bibitem{Lees:2012tva}
{BaBar collaboration}, J.~P. Lees {\em et~al.},
  \ifthenelse{\boolean{articletitles}}{{\it {Measurement of branching fractions
  and rate asymmetries in the rare decays $B \to K^{(*)} l^+ l^-$}},
  }{}\href{http://dx.doi.org/10.1103/PhysRevD.86.032012}{Phys.\ Rev.\  {\bf
  D86} (2012) 032012}, \href{http://arxiv.org/abs/1204.3933}{{\tt
  arXiv:1204.3933}}\relax
\mciteBstWouldAddEndPuncttrue
\mciteSetBstMidEndSepPunct{\mcitedefaultmidpunct}
{\mcitedefaultendpunct}{\mcitedefaultseppunct}\relax
\EndOfBibitem
\bibitem{Wei:2009zv}
Belle collaboration, J.-T. Wei {\em et~al.},
  \ifthenelse{\boolean{articletitles}}{{\it {Measurement of the differential
  branching fraction and forward-backward asymmetry for $B\rightarrow K^{(*)}
  \ellell$}}, }{}\href{http://dx.doi.org/10.1103/PhysRevLett.103.171801}{Phys.\
  Rev.\ Lett.\  {\bf 103} (2009) 171801},
  \href{http://arxiv.org/abs/0904.0770}{{\tt arXiv:0904.0770}}\relax
\mciteBstWouldAddEndPuncttrue
\mciteSetBstMidEndSepPunct{\mcitedefaultmidpunct}
{\mcitedefaultendpunct}{\mcitedefaultseppunct}\relax
\EndOfBibitem
\bibitem{LHCb-PAPER-2013-039}
LHCb collaboration, R.~Aaij {\em et~al.},
  \ifthenelse{\boolean{articletitles}}{{\it {Observation of a resonance in
  $B^+\to K^+\mu^+\mu^-$ decays at low recoil}},
  }{}\href{http://dx.doi.org/10.1103/PhysRevLett.111.112003}{Phys.\ Rev.\
  Lett.\  {\bf 111} (2013) 112003}, \href{http://arxiv.org/abs/1307.7595}{{\tt
  arXiv:1307.7595}}\relax
\mciteBstWouldAddEndPuncttrue
\mciteSetBstMidEndSepPunct{\mcitedefaultmidpunct}
{\mcitedefaultendpunct}{\mcitedefaultseppunct}\relax
\EndOfBibitem
\bibitem{PDG2012}
Particle Data Group, J.~Beringer {\em et~al.},
  \ifthenelse{\boolean{articletitles}}{{\it {\href{http://pdg.lbl.gov/}{Review
  of particle physics}}},
  }{}\href{http://dx.doi.org/10.1103/PhysRevD.86.010001}{Phys.\ Rev.\  {\bf
  D86} (2012) 010001}, {and 2013 partial update for the 2014 edition}\relax
\mciteBstWouldAddEndPuncttrue
\mciteSetBstMidEndSepPunct{\mcitedefaultmidpunct}
{\mcitedefaultendpunct}{\mcitedefaultseppunct}\relax
\EndOfBibitem
\bibitem{Alves:2008zz}
LHCb collaboration, A.~A. Alves~Jr. {\em et~al.},
  \ifthenelse{\boolean{articletitles}}{{\it {The \lhcb detector at the LHC}},
  }{}\href{http://dx.doi.org/10.1088/1748-0221/3/08/S08005}{JINST {\bf 3}
  (2008) S08005}\relax
\mciteBstWouldAddEndPuncttrue
\mciteSetBstMidEndSepPunct{\mcitedefaultmidpunct}
{\mcitedefaultendpunct}{\mcitedefaultseppunct}\relax
\EndOfBibitem
\bibitem{Sjostrand:2007gs}
T.~Sj\"{o}strand, S.~Mrenna, and P.~Skands,
  \ifthenelse{\boolean{articletitles}}{{\it {A brief introduction to PYTHIA
  8.1}}, }{}\href{http://dx.doi.org/10.1016/j.cpc.2008.01.036}{Comput.\ Phys.\
  Commun.\  {\bf 178} (2008) 852}, \href{http://arxiv.org/abs/0710.3820}{{\tt
  arXiv:0710.3820}}\relax
\mciteBstWouldAddEndPuncttrue
\mciteSetBstMidEndSepPunct{\mcitedefaultmidpunct}
{\mcitedefaultendpunct}{\mcitedefaultseppunct}\relax
\EndOfBibitem
\bibitem{LHCb-PROC-2010-056}
I.~Belyaev {\em et~al.}, \ifthenelse{\boolean{articletitles}}{{\it {Handling of
  the generation of primary events in \gauss, the \lhcb simulation framework}},
  }{}\href{http://dx.doi.org/10.1109/NSSMIC.2010.5873949}{Nuclear Science
  Symposium Conference Record (NSS/MIC) {\bf IEEE} (2010) 1155}\relax
\mciteBstWouldAddEndPuncttrue
\mciteSetBstMidEndSepPunct{\mcitedefaultmidpunct}
{\mcitedefaultendpunct}{\mcitedefaultseppunct}\relax
\EndOfBibitem
\bibitem{Lange:2001uf}
D.~J. Lange, \ifthenelse{\boolean{articletitles}}{{\it {The EvtGen particle
  decay simulation package}},
  }{}\href{http://dx.doi.org/10.1016/S0168-9002(01)00089-4}{Nucl.\ Instrum.\
  Meth.\  {\bf A462} (2001) 152}\relax
\mciteBstWouldAddEndPuncttrue
\mciteSetBstMidEndSepPunct{\mcitedefaultmidpunct}
{\mcitedefaultendpunct}{\mcitedefaultseppunct}\relax
\EndOfBibitem
\bibitem{Golonka:2005pn}
P.~Golonka and Z.~Was, \ifthenelse{\boolean{articletitles}}{{\it {PHOTOS Monte
  Carlo: a precision tool for QED corrections in $Z$ and $W$ decays}},
  }{}\href{http://dx.doi.org/10.1140/epjc/s2005-02396-4}{Eur.\ Phys.\ J.\  {\bf
  C45} (2006) 97}, \href{http://arxiv.org/abs/hep-ph/0506026}{{\tt
  arXiv:hep-ph/0506026}}\relax
\mciteBstWouldAddEndPuncttrue
\mciteSetBstMidEndSepPunct{\mcitedefaultmidpunct}
{\mcitedefaultendpunct}{\mcitedefaultseppunct}\relax
\EndOfBibitem
\bibitem{Allison:2006ve}
Geant4 collaboration, J.~Allison {\em et~al.},
  \ifthenelse{\boolean{articletitles}}{{\it {Geant4 developments and
  applications}}, }{}\href{http://dx.doi.org/10.1109/TNS.2006.869826}{IEEE
  Trans.\ Nucl.\ Sci.\  {\bf 53} (2006) 270}\relax
\mciteBstWouldAddEndPuncttrue
\mciteSetBstMidEndSepPunct{\mcitedefaultmidpunct}
{\mcitedefaultendpunct}{\mcitedefaultseppunct}\relax
\EndOfBibitem
\bibitem{Agostinelli:2002hh}
Geant4 collaboration, S.~Agostinelli {\em et~al.},
  \ifthenelse{\boolean{articletitles}}{{\it {Geant4: a simulation toolkit}},
  }{}\href{http://dx.doi.org/10.1016/S0168-9002(03)01368-8}{Nucl.\ Instrum.\
  Meth.\  {\bf A506} (2003) 250}\relax
\mciteBstWouldAddEndPuncttrue
\mciteSetBstMidEndSepPunct{\mcitedefaultmidpunct}
{\mcitedefaultendpunct}{\mcitedefaultseppunct}\relax
\EndOfBibitem
\bibitem{LHCb-PROC-2011-006}
M.~Clemencic {\em et~al.}, \ifthenelse{\boolean{articletitles}}{{\it {The \lhcb
  simulation application, \gauss: design, evolution and experience}},
  }{}\href{http://dx.doi.org/10.1088/1742-6596/331/3/032023}{{J.\ Phys.\ Conf.\
  Ser.\ } {\bf 331} (2011) 032023}\relax
\mciteBstWouldAddEndPuncttrue
\mciteSetBstMidEndSepPunct{\mcitedefaultmidpunct}
{\mcitedefaultendpunct}{\mcitedefaultseppunct}\relax
\EndOfBibitem
\bibitem{BBDT}
V.~V. Gligorov and M.~Williams, \ifthenelse{\boolean{articletitles}}{{\it
  {Efficient, reliable and fast high-level triggering using a bonsai boosted
  decision tree}},
  }{}\href{http://dx.doi.org/10.1088/1748-0221/8/02/P02013}{JINST {\bf 8}
  (2013) P02013}, \href{http://arxiv.org/abs/1210.6861}{{\tt
  arXiv:1210.6861}}\relax
\mciteBstWouldAddEndPuncttrue
\mciteSetBstMidEndSepPunct{\mcitedefaultmidpunct}
{\mcitedefaultendpunct}{\mcitedefaultseppunct}\relax
\EndOfBibitem
\bibitem{LHCB-DP-2013-001}
F.~Archilli {\em et~al.}, \ifthenelse{\boolean{articletitles}}{{\it
  {Performance of the muon identification at LHCb}},
  }{}\href{http://dx.doi.org/10.1088/1748-0221/8/10/P10020}{JINST {\bf 8}
  (2013) P10020}, \href{http://arxiv.org/abs/1306.0249}{{\tt
  arXiv:1306.0249}}\relax
\mciteBstWouldAddEndPuncttrue
\mciteSetBstMidEndSepPunct{\mcitedefaultmidpunct}
{\mcitedefaultendpunct}{\mcitedefaultseppunct}\relax
\EndOfBibitem
\bibitem{Breiman}
L.~Breiman, J.~H. Friedman, R.~A. Olshen, and C.~J. Stone, {\em Classification
  and regression trees}, Wadsworth international group, Belmont, California,
  USA, 1984\relax
\mciteBstWouldAddEndPuncttrue
\mciteSetBstMidEndSepPunct{\mcitedefaultmidpunct}
{\mcitedefaultendpunct}{\mcitedefaultseppunct}\relax
\EndOfBibitem
\bibitem{Roe}
B.~P. Roe {\em et~al.}, \ifthenelse{\boolean{articletitles}}{{\it {Boosted
  decision trees as an alternative to artificial neural networks for particle
  identification}},
  }{}\href{http://dx.doi.org/10.1016/j.nima.2004.12.018}{Nucl.\ Instrum.\
  Meth.\  {\bf A543} (2005) 577},
  \href{http://arxiv.org/abs/physics/0408124}{{\tt
  arXiv:physics/0408124}}\relax
\mciteBstWouldAddEndPuncttrue
\mciteSetBstMidEndSepPunct{\mcitedefaultmidpunct}
{\mcitedefaultendpunct}{\mcitedefaultseppunct}\relax
\EndOfBibitem
\bibitem{Skwarnicki:1986xj}
T.~Skwarnicki, {\em {A study of the radiative cascade transitions between the
  Upsilon-prime and Upsilon resonances}}, PhD thesis, Institute of Nuclear
  Physics, Krakow, 1986,
  {\href{http://inspirehep.net/record/230779/files/230779.pdf}{DESY-F31-86-02}%
}\relax
\mciteBstWouldAddEndPuncttrue
\mciteSetBstMidEndSepPunct{\mcitedefaultmidpunct}
{\mcitedefaultendpunct}{\mcitedefaultseppunct}\relax
\EndOfBibitem
\bibitem{LHCb-PAPER-2013-005}
LHCb collaboration, R.~Aaij {\em et~al.},
  \ifthenelse{\boolean{articletitles}}{{\it {Measurement of the $B^0 \to
  K^{*0}e^+e^-$ branching fraction at low dilepton mass}},
  }{}\href{http://dx.doi.org/10.1007/JHEP05(2013)159}{JHEP {\bf 05} (2013)
  159}, \href{http://arxiv.org/abs/1304.3035}{{\tt arXiv:1304.3035}}\relax
\mciteBstWouldAddEndPuncttrue
\mciteSetBstMidEndSepPunct{\mcitedefaultmidpunct}
{\mcitedefaultendpunct}{\mcitedefaultseppunct}\relax
\EndOfBibitem
\bibitem{Bobeth:2010wg}
C.~Bobeth, G.~Hiller, and D.~van Dyk, \ifthenelse{\boolean{articletitles}}{{\it
  {The benefits of $\bar{B} \to \bar{K}^* l^+ l^-$ decays at low recoil}},
  }{}\href{http://dx.doi.org/10.1007/JHEP07(2010)098}{JHEP {\bf 07} (2010)
  098}, \href{http://arxiv.org/abs/1006.5013}{{\tt arXiv:1006.5013}}\relax
\mciteBstWouldAddEndPuncttrue
\mciteSetBstMidEndSepPunct{\mcitedefaultmidpunct}
{\mcitedefaultendpunct}{\mcitedefaultseppunct}\relax
\EndOfBibitem
\bibitem{LHCb-PAPER-2011-035}
LHCb collaboration, R.~Aaij {\em et~al.},
  \ifthenelse{\boolean{articletitles}}{{\it {Measurement of $b$-hadron
  masses}}, }{}\href{http://dx.doi.org/10.1016/j.physletb.2012.01.058}{Phys.\
  Lett.\  {\bf B708} (2012) 241}, \href{http://arxiv.org/abs/1112.4896}{{\tt
  arXiv:1112.4896}}\relax
\mciteBstWouldAddEndPuncttrue
\mciteSetBstMidEndSepPunct{\mcitedefaultmidpunct}
{\mcitedefaultendpunct}{\mcitedefaultseppunct}\relax
\EndOfBibitem
\bibitem{LHCb-PAPER-2012-048}
LHCb collaboration, R.~Aaij {\em et~al.},
  \ifthenelse{\boolean{articletitles}}{{\it {Measurements of the $\Lambda_b^0$,
  $\Xi_b^-$ and $\Omega_b^-$ baryon masses}},
  }{}\href{http://dx.doi.org/10.1103/PhysRevLett.110.182001}{Phys.\ Rev.\
  Lett.\  {\bf 110} (2013) 182001}, \href{http://arxiv.org/abs/1302.1072}{{\tt
  arXiv:1302.1072}}\relax
\mciteBstWouldAddEndPuncttrue
\mciteSetBstMidEndSepPunct{\mcitedefaultmidpunct}
{\mcitedefaultendpunct}{\mcitedefaultseppunct}\relax
\EndOfBibitem
\bibitem{LHCb-PAPER-2012-010}
LHCb collaboration, R.~Aaij {\em et~al.},
  \ifthenelse{\boolean{articletitles}}{{\it {Measurement of relative branching
  fractions of $B$ decays to $\psi(2S)$ and $J/\psi$ mesons}},
  }{}\href{http://dx.doi.org/10.1140/epjc/s10052-012-2118-7}{Eur.\ Phys.\ J.\
  {\bf C72} (2012) 2118}, \href{http://arxiv.org/abs/1205.0918}{{\tt
  arXiv:1205.0918}}\relax
\mciteBstWouldAddEndPuncttrue
\mciteSetBstMidEndSepPunct{\mcitedefaultmidpunct}
{\mcitedefaultendpunct}{\mcitedefaultseppunct}\relax
\EndOfBibitem
\bibitem{LHCb-PAPER-2013-024}
LHCb collaboration, R.~Aaij {\em et~al.},
  \ifthenelse{\boolean{articletitles}}{{\it {Observation of
  $B^0_s\to\chi_{c1}\phi$ decay and study of $B^0\to\chi_{c1,2}K^{*0}$
  decays}}, }{}\href{http://dx.doi.org/10.1016/j.nuclphysb.2013.06.005}{Nucl.\
  Phys.\  {\bf B874} (2013) 663}, \href{http://arxiv.org/abs/1305.6511}{{\tt
  arXiv:1305.6511}}\relax
\mciteBstWouldAddEndPuncttrue
\mciteSetBstMidEndSepPunct{\mcitedefaultmidpunct}
{\mcitedefaultendpunct}{\mcitedefaultseppunct}\relax
\EndOfBibitem
\end{mcitethebibliography}
\ifx\mcitethebibliography\mciteundefinedmacro
\PackageError{LHCb.bst}{mciteplus.sty has not been loaded}
{This bibstyle requires the use of the mciteplus package.}\fi
\providecommand{\href}[2]{#2}



\end{document}